# Hydrological post-processing for predicting extreme quantiles

Hristos Tyralis[1,2,3,*], Georgia Papacharalampous[4]

[1]Construction Agency, Hellenic Air Force, Mesogion Avenue 227–231, 15 561 Cholargos, Greece (montchrister@gmail.com, https://orcid.org/0000-0002-8932-4997)

[2]Department of Topography, School of Rural, Surveying and Geoinformatics Engineering, National Technical University of Athens, Iroon Polytechniou 5, 157 80 Zografou, Greece (montchrister@gmail.com, https://orcid.org/0000-0002-8932-4997)

[3]Department of Water Resources and Environmental Engineering, School of Civil Engineering, National Technical University of Athens, Iroon Polytechniou 5, 157 80 Zografou, Greece (hristos@itia.ntua.gr, https://orcid.org/0000-0002-8932-4997)

[4]Department of Topography, School of Rural, Surveying and Geoinformatics Engineering, National Technical University of Athens, Iroon Polytechniou 5, 157 80 Zografou, Greece (papacharalampous.georgia@gmail.com, https://orcid.org/0000-0001-5446-954X)

*Corresponding author

**Abstract**: Hydrological post-processing using quantile regression algorithms constitutes a prime means of estimating the uncertainty of hydrological predictions. Nonetheless, conventional large-sample theory for quantile regression does not apply sufficiently far in the tails of the probability distribution of the dependent variable. To overcome this limitation that could be crucial when the interest lies on flood events, hydrological post-processing through extremal quantile regression is introduced here for estimating the extreme quantiles of hydrological model's responses. In summary, the new hydrological post-processing method exploits properties of the Hill's estimator from the extreme value theory to extrapolate quantile regression's predictions to high quantiles. As a proof of concept, the new method is here tested in post-processing daily streamflow simulations provided by three process-based hydrological models for 180 basins in the contiguous United States (CONUS) and is further compared to conventional quantile regression. With this large-scale comparison, it is demonstrated that hydrological post-processing using conventional quantile regression severely underestimates high quantiles (at the quantile level 0.9999) compared to hydrological post-processing using extremal quantile regression, although both methods are equivalent at lower quantiles (at the quantile level 0.9700). Moreover, it is shown that, in the same context, extremal quantile regression estimates the high predictive quantiles with efficiency that is, on average, equivalent in the large-sample study for the three process-based hydrological models.

**Keywords**: big data; computational hydrology; extreme value theory; quantile regression; rainfall-runoff model; uncertainty estimation

# 1. Introduction

One purpose of hydrological models is to predict the responses of hydrological systems. Predictions of models are most informative and complete when they are probabilistic in nature, taking the form of probability distributions (Gneiting and Raftery 2007), although it is a frequent practice to utilize hydrological models in a deterministic manner, in the "best guess" sense which requires the predictions to be close to their corresponding observations (Farmer and Vogel 2016). Indeed, estimating the uncertainty of hydrological predictions is receiving increasing attention nowadays (Beven and Binley 1992, Todini 2007, Montanari 2011), while disentangling and reducing this same uncertainty has been identified as the 20[th] unsolved problem in hydrology (Blöschl et al. 2019).

A class of methods for estimating the uncertainty of hydrological predictions includes joint approaches, in which the residual error model and hydrological model parameters are inferred simultaneously. This is possible through Monte Carlo or Bayesian-based methods (Beven and Binley 1992, Evin et al. 2014), or by adapting the loss function of the models aiming to directly estimate quantiles or expectiles of the probability distribution (Tyralis and Papacharalampous 2021, Tyralis et al. 2023). The second class of methods includes post-processor (two-stage) approaches (Montanari and Koutsoyiannis 2012, Sikorska et al. 2015, Li et al. 2017, Biondi and Todini 2018, Papacharalampous et al. 2019, Tyralis et al. 2019a, Papacharalampous et al. 2020a, 2020b, Sikorska-Senoner and Quilty 2021, Li et al. 2021a, Koutsoyiannis and Montanari 2022, Quilty et al. 2022).

The first stage of the post-processor approaches consists in estimating a functional (Gneiting 2011) of the probability distribution of the hydrological model's response that is close to the observations (e.g. the mean of the model's response estimated by calibrating the process-based hydrological model with a squared error type function,

Nash and Sutcliffe 1970), while their second stage consists in modelling the residual errors using a statistical or machine learning algorithm that can estimate the probability distribution of the hydrological model response. One post-processor class includes ensemble models (Quilty et al. 2019, Quilty and Adamowski 2020). Another post-processor class is based on the utilization of quantile regression algorithms, either the linear–in-parameters quantile regression (Dogulu et al. 2015, Papacharalampous et al. 2020a, 2020b) or quantile-based machine learning algorithms (Papacharalampous et al. 2019, Tyralis et al. 2019a). A review on probabilistic hydrological post-processing methods with machine learning algorithms can be found in Papacharalampous and Tyralis (2022).

Prediction of extreme events is of interest in hydrology (O'Gorman 2015, Donat et al. 2016, Pfahl et al. 2017), because of the severity of socio-economic losses when those events take place (Vogel et al. 2018). When flood events are of interest (Bogner et al. 2016), a common practice in the field of hydrological modelling is to adapt a process-based hydrological model for predicting high flows. Frequently, the model is calibrated and assessed using squared error type metrics (Nash and Sutcliffe 1970). This practice is consistent for estimating the mean of the probability distribution of predicted high flows (Gneiting 2011). On the other hand, this practice does not allow probabilistic predictions.

Prediction of floods naturally should be associated with extreme quantiles of the predictive probability distribution of the hydrological model's response. Therefore, from a statistical point of view the problem of predicting flood events could be alternatively formulated as a predictive probability distribution estimation, where one is interested in high (or extreme quantiles) of the probability distribution of the hydrological model's response (but not the mean of the probability distribution of the hydrological model's response). By limiting the scope of the problem to specific quantiles of the predictive probability distribution, instead of the full probability distribution, one can calibrate the hydrological model with a quantile or expectile loss function, targeting quantiles or expectiles at a pre-specified level (e.g. 0.99-level) of interest (e.g. as done in Tyralis and Papacharalampous 2021, Tyralis et al. 2023), or one can post-process hydrological simulations using conventional quantile regression (Papacharalampous and Tyralis 2022). However, conventional quantile regression is not suitable for predicting extreme quantiles of the probability distribution, especially when the hydrological time series are short because conventional large sample theory for quantile regression does not apply sufficiently far in the tails (Chernozhukov 2005).

In some cases that deviate from standard practice of targeting the mean of the probability distribution of predicted high flows, Krzysztofowicz (2014) has estimated the predictive exceedance function of the maximum river stages. In another relevant case, Han and Coulibaly (2019) have used ensemble weather forecasts. A natural alternative to these approaches that is based on the extreme value theory has been explored by Curceac et al. (2020, 2021) who adjusted for conditional bias when simulating conditional extremes. Another approach is based on Bayesian causal modelling where the causal Bayesian model is trained with historical extreme rainfall records (Molina et al. 2022).

To solve the problem of predicting extreme quantiles of the probability distribution of hydrological model's responses in the overall post-processing paradigm, which is well-established for uncertainty estimation in the field (see the literature summary above) but has not been formulated yet in the endeavour of predicting extreme quantiles, a new post-processing method based on conditional extreme quantile regression, and more precisely on the algorithm proposed by Wang et al. (2012) and Wang and Li (2013) is proposed here. This algorithm elaborates on the extreme value theory for extrapolating at higher quantiles predictions that have been previously issued using quantile regression. The method is applied to a large dataset comprising information from basins in the contiguous United States (CONUS) for post-processing simulations by three Génie Rural (GR) daily lumped hydrological models (Coron et al. 2017), namely the GR4J, GR5J and GR6J.

The remainder of the paper is structured as follows. Section 2 presents the methods, as well as the large multi-site dataset to which the methods were applied, followed by an outline of the proposed methodology. The results are presented and discussed in Sections 3 and 4, respectively. The provided discussions focus on the theoretical properties of the herein proposed method in contrast to methods that have been previously proposed and implemented in the literature. The paper concludes with Section 5.

## 2. Methods and data

Here, the concepts introduced in the manuscript together with the methods and the application's data are presented. For information on software implementation, see Appendix A.

## 2.1 Hydrological modelling and post-processing

Basic information on the hydrological models utilized in the study and the concept of post-processing is summarized here.

### 2.1.1 Hydrological models

The methods were applied to simulations provided by the GR4J (Perrin et al. 2003), GR5J and GR6J (Pushpalatha et al. 2011) daily lumped hydrological models. A detailed description of these hydrological models is out of scope of the present work, which focuses on post-processing techniques (that could be applied to the simulations provided by any hydrological model). Moreover, the GR hydrological models are widely applied and well established in the hydrological community. The number $x$ in the name GR$x$J denotes the number of parameters in the corresponding model, with more parameters increasing the flexibility and improving the modelling properties. Compared to the GR4J model, the GR5J model can better represent inter-catchment water exchanges, while the GR6J model can provide improved predictions of low flows (Coron et al. 2017). Michel's (1991) optimization algorithm was used for calibrating the hydrological models.

### 2.1.2 Hydrological post-processing

Hydrological post-processing was performed by regressing the observed daily streamflow at the time $t$ (dependent variable) to the simulated daily streamflow at the same time $t$ (predictor variable). Linear-in-parameters quantile regression (hereinafter simply referred to as "quantile regression") (Koenker and Bassett Jr 1978) and extremal quantile regression (Wang et al. 2012, Wang and Li 2013) were utilized as regression algorithms. Their description follows.

## 2.2 Quantile regression

Observations will be notated with lowercase letters, while respective random variables will be notated by respective underlined lowercase letters. Let $\underline{y}$ be a random variable with cumulative distribution function $F_{\underline{y}}$ defined by:

$$F_{\underline{y}}(y) := P(\underline{y} \leq y) \qquad (1)$$

Then, $Q_{\underline{y}}(\tau)$ defined by:

$$Q_{\underline{y}}(\tau) := \inf\{y : F_{\underline{y}}(y) \geq \tau\} \qquad (2)$$

is referred to as the $\tau^{\text{th}}$ quantile of $\underline{y}$, where inf denotes the infimum of a set of real numbers. A special case of eq. (2) is the median of $y$, $Q_y(1/2)$.

Let $F_{y|x}$ be the distribution of the random variable $y$ given the $p$-dimensional vector $\boldsymbol{x}$:

$$F_{y|x}(y|\boldsymbol{x}) := P(y \leq y|\boldsymbol{x}) \tag{3}$$

Then, $Q_{y|x}(\tau|\boldsymbol{x})$ defined by:

$$Q_{y|x}(\tau|\boldsymbol{x}) := \inf\{y: F_y(y|\boldsymbol{x}) \geq \tau\} \tag{4}$$

is referred to as the $\tau^{\text{th}}$ quantile of $y$ conditional on $\boldsymbol{x}$. $Q_{y|x}(\tau|\boldsymbol{x})$ can be estimated by a linear model defined by:

$$Q_{y|x}(\tau|\boldsymbol{x}) := \alpha(\tau) + \boldsymbol{x}^{\text{T}} \boldsymbol{\beta}(\tau), \ 0 < \tau < 1 \tag{5}$$

where $\alpha(\tau)$ is a scalar parameter and $\boldsymbol{\beta}(\tau)$ is a $p$-dimensional parameter vector. One can estimate $Q_{y|x}(\tau|\boldsymbol{x})$ by first estimating the parameters $\alpha(\tau)$ and $\boldsymbol{\beta}(\tau)$ of eq. (5).

Let $\{(y_i, \boldsymbol{x}_i), i = 1, \ldots, n\}$ be a random sample from the vector $(y, \boldsymbol{x})$ and

$$\rho_{1/2}(u) = |u|/2 \tag{6}$$

where $\rho_{1/2}(u)$ is half the absolute error function. By minimizing the mean absolute error (MAE) $\Sigma_{i=1}^{n} 2 \rho_{1/2}(y_i - \alpha(1/2) - \boldsymbol{x}_i^{\text{T}} \boldsymbol{\beta}(1/2))/n$, one can estimate the parameters $\alpha(1/2)$ and $\boldsymbol{\beta}(1/2)$ that correspond to the median $Q_{y|x}(1/2|\boldsymbol{x})$ of $y|\boldsymbol{x}$.

The absolute error function is generalized by the quantile loss function defined by eq. (7):

$$\rho_\tau(u) := u \ (\tau - \mathbb{I}(u \leq 0)) \tag{7}$$

Here $\mathbb{I}(\cdot)$ denotes the indicator function and $\tau$ is the quantile level of interest. For $\tau = 1/2$, eq. (7) reduces to eq. (6).

To estimate $\alpha(\tau)$ and $\boldsymbol{\beta}(\tau)$ for $\tau \in (0, 1)$, one should minimize the average quantile score $(1/n) \Sigma_{i=1}^{n} \rho_\tau(y_i - \alpha(\tau) - \boldsymbol{x}_i^{\text{T}} \boldsymbol{\beta}(\tau))$, which is the core idea of quantile regression elaborated by Koenker and Bassett Jr (1978). The quantile loss function, defined by eq. (7), is positive and negatively oriented, i.e. the objective is to minimize it, and equals to 0, when $u = 0$.

## 2.3 Extremal quantile regression

Here, the algorithm that can support the estimation of extreme conditional quantiles (Wang et al. 2012, Wang and Li 2013) is presented. Quantile regression is unstable when estimating quantiles at the tails due to data sparseness. Therefore, some adaptations are needed to estimate $Q_{y|x}(\tau_n|\boldsymbol{x})$, where $\tau_n \to 1$ as $n \to \infty$. An assumption of the algorithm is that $F_{y|x}(\cdot|\boldsymbol{x})$ is in the domain of attraction $D(G_\gamma)$ of an extreme value distribution $G_\gamma(\cdot)$, i.e.

$F_{\underline{y}}(\cdot|\boldsymbol{x}) \in D(G_\gamma)$ where $\gamma > 0$ is the extreme value index (EVI) (Fisher and Tippett 1928, Gnedenko 1943). The aim is to estimate the following linear quantile regression model:

$$Q_{\underline{y}|\boldsymbol{x}}(\tau|\boldsymbol{x}) := \alpha(\tau) + \boldsymbol{x}^T \boldsymbol{\beta}(\tau), \forall \tau \in [\tau_c, 1] \tag{8}$$

Define a sequence of quantile levels $\tau_c < \tau_{n-k} < \tau_{n-k+1} < \ldots < \tau_m \in (0, 1)$, where $m = n - [n^v]$, with $[\cdot]$ denoting the integer part of a number, $v > 0$ is some small constant, and $\tau_j = j/(n+1)$. Some further assumptions are $k = k_n \to \infty$, $k/n \to 0$ as $n \to \infty$, and $v > 0$ such that $n^v < k$. For each $j = n - k, \ldots, m$, $\hat{\alpha}(\tau_j)$ and $\widehat{\boldsymbol{\beta}}(\tau_j)$ are defined by:

$$(\hat{\alpha}(\tau_j), \widehat{\boldsymbol{\beta}}(\tau_j)^T)^T = \arg\min_{\alpha, \boldsymbol{\beta}} \sum_{i=1}^{n} \rho_{\tau_j}(y_i - \alpha - \boldsymbol{x}_i^T \boldsymbol{\beta}) \tag{9}$$

For a given $\boldsymbol{x}$, define

$$q_j := \hat{\alpha}(\tau_j) + \boldsymbol{x}^T \widehat{\boldsymbol{\beta}}(\tau_j), j = n - k, \ldots, m \tag{10}$$

The parameters $q_j$ can be roughly regarded as the upper order statistics of a sample from $F_{\underline{y}|\boldsymbol{x}}(\cdot|\boldsymbol{x})$. Consequently, under the assumption $F_{\underline{y}}(\cdot|\boldsymbol{x}) \in D(G_\gamma)$, where $\gamma > 0$, $\gamma$ can be estimated by:

$$\hat{\gamma}(\boldsymbol{x}) = (1/(k - [n^v])) \sum_{j=[n^v]}^{k} \log(q_{n-j}/q_{n-k}) \tag{11}$$

where $\hat{\gamma}$ can be viewed as the Hill's (1975) estimator. Consequently, $\hat{Q}_{\underline{y}}(\tau_n|\boldsymbol{x})$ can be estimated by an adaptation of the Weissman's (1978) estimator:

$$\hat{Q}_{\underline{y}|\boldsymbol{x}}(\tau_n|\boldsymbol{x}) := ((1 - \tau_{n-k})/(1 - \tau_n))^{\hat{\gamma}} q_{n-k} \tag{12}$$

Eq. (12) is essentially an extrapolation of intermediate quantile estimates to the far tails. Following numerical experiments by Wang et al. (2012), $v = 0.1$ is set. The Hill's estimator has several favourable properties, e.g. it is a consistent estimator of the EVI. On the other hand, a problem is that for different $k$, different Hill's estimators are obtained. A summary of Hills's estimator properties can be found in Beirlant et al. (2004, Section 4.2). The parameter $k$ is selected using the procedure proposed by Wang and Li (2013). Small values of $k$ lead to large variance, while large values of $k$ lead to more bias in the estimation of $\gamma(\boldsymbol{x})$. Furthermore, the pooled estimator of $\gamma$ is used (Wang et al. 2012, Section 2.2.3) instead of those varying conditional on $\boldsymbol{x}$.

$$\hat{\gamma}_{\text{pool}} := (1/n) \sum_{j=1}^{n} \hat{\gamma}(\boldsymbol{x}_j) \tag{13}$$

At this point it is relevant to mention that modelling heavy tails in quantile regression settings is also possible through other techniques that also address the frequently met problem of quantile crossing. For instance, this is the case in Adlouni and Baldé (2019) who implemented quantile regression in a Bayesian setting using a working likelihood (in particular the asymmetric Laplace distribution). Different types of quantile regression for predicting extremes quantiles could be the subject of further research.

## 2.4 Data

Daily time series of streamflow, precipitation, and minimum and maximum temperatures were extracted from the CAMELS dataset (Addor et al. 2017a, b, Newman et al. 2014, 2015, 2017, Thornton et al. 2014). These time series extend in the 7-year period of 2007–2013 and originate from 180 small- to medium-sized basins located in the contiguous United States (CONUS), as shown in to Figure 1. The short periods were selected to show the benefits of the method compared to conventional quantile regression, while the case is relevant given the limited availability of data mostly in developing countries. The daily mean temperature was estimated by averaging the minimum and maximum daily temperatures. Oudin's formula (Oudin et al. 2005) was applied to the daily mean temperature for estimating the daily potential evapotranspiration.

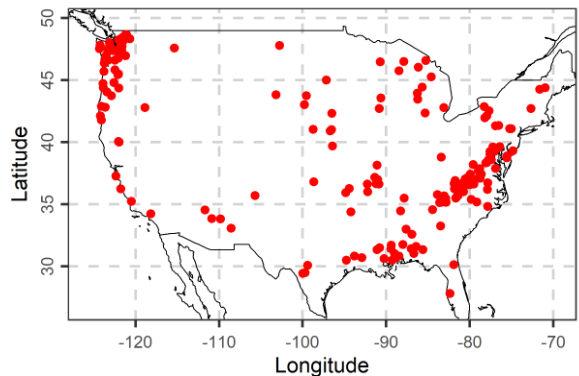

Figure 1. The 180 basins examined in the study.

## 2.5 Methodology outline

Here, an outline of the proposed methodology is presented, while specific components have been presented in previous Sections. The framework is applicable to a single basin and a single hydrological model. The procedure is repeated for every hydrological model (GR4J, GR5J, GR6J, see Section 2.1.1) to every basin of the dataset and summary results for the set of basins are presented in Section 3. Let $T_1$ = {2007-01-01, …, 2007-12-31}, $T_2$ = {2008-01-01, …, 2009-12-31}, $T_3$ = {2010-01-01, …, 2011-12-31} and $T_4$ = {2012-01-01, …, 2013-12-31}. The inputs to the hydrological model are daily precipitation and potential evapotranspiration, while the output is daily streamflow; see Section 2.4. Let $y_i$ be the observed streamflow at the day $i \in \{T_1, T_2, T_3, T_4\}$ at a random basin. The procedure of the framework follows:

Step 1: Warm-up the hydrological model in the period $T_1$.

Step 2: Calibrate the hydrological model in the period $T_2$ by using the squared error function and the optimization algorithm by Michel ([1991](#)).

Step 3: Simulate streamflow by using the hydrological model in the period $\{T_3, T_4\}$. Let $x_i$, be the simulated streamflow at the day $i \in \{T_3, T_4\}$.

Step 4: Apply quantile regression and extremal quantile regression at the levels $\tau \in \{0.9700, 0.9990, 0.9999\}$ in the period $T_3$ by using $y_i$, where $i \in T_3$, as the dependent variable and $x_i$, where $i \in T_3$, as the predictor variable.

Step 5: Predict quantiles of the probability distribution of model's predicted streamflow at the levels $\tau \in \{0.9700, 0.9990, 0.9999\}$ by using quantile regression and extremal quantile regression in period $T_4$ with $x_i$, where $i \in T_4$, as the predictor variable.

The final product consists of 2 (number of post-processors) × 3 (number of quantile levels) × 180 (number of basins) = 1 080 simulated streamflow series (here, simulation refers to quantiles of the probability distribution of model's predicted streamflow) at the daily resolution and in the period $T_4$, in which the post-processors are tested. Each simulated streamflow series is composed by 731 values, leading to a total of 1 080 × 731 = 789 480 values in the period $T_4$. The results presented in the following refer to the test period $T_4$.

## 3. Results

A representative example of post-processed simulations is provided in [Figure 2](#), where estimated quantiles of post-processed model's response at the levels $\tau \in \{0.9700, 0.9990, 0.9999\}$ are presented for an arbitrary basin and for both regular and extremal quantile regression. [Figure 2b](#) zooms for illustration purposes on a one-month period that consists a subset of the two-year period presented in [Figure 2a](#). Recalling that post-processing is used to turn point predictions to probabilistic, a first indication that both post-processing methods are reliable is that the estimated quantiles are larger than the observations. Both methods are almost equivalent in estimating quantiles at the levels 0.9700 and 0.9990. However, the quantiles estimated by quantile regression at the level 0.9999 are almost equivalent with the quantiles at the level 0.9990. On the contrary, extremal quantile regression extrapolated beyond the level 0.9990, as for it the estimated quantiles at the level 0.9999 are considerably larger. This illustration is typical of the adverse properties of quantile regression at predicting quantiles at the far tails.

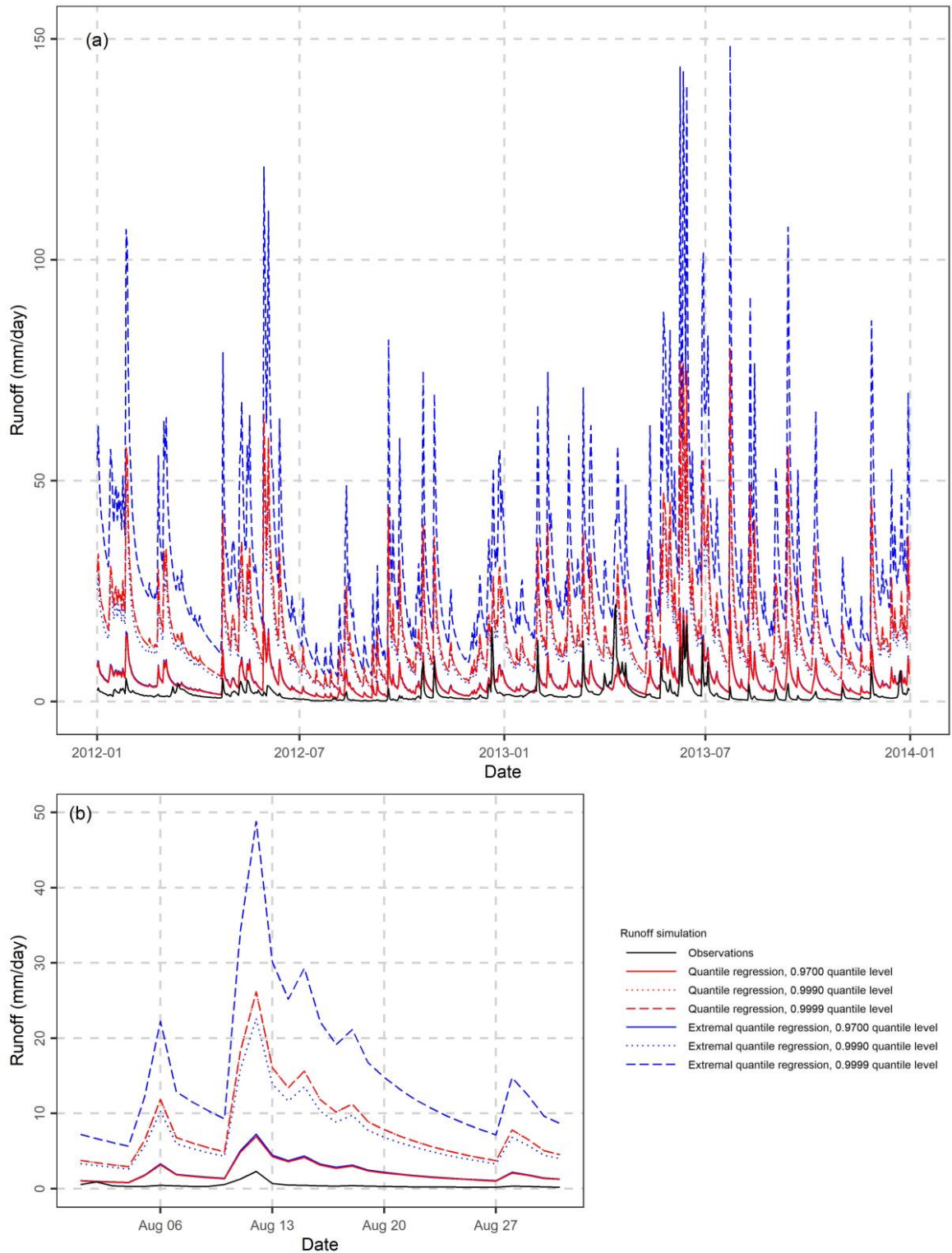

Figure 2. Illustration of observed streamflow and model's response quantiles estimated by post-processing the simulations of the GR4J model by using quantile and extremal quantile regression at the levels $a \in \{0.9700, 0.9990, 0.9999\}$ for a two-year period (top) and a one-month subset of the same period (bottom) at an arbitrary basin.

Practically, when referring e.g. to level 0.9999, and given daily rainfall and streamflow observations for 10 000 days (approximately 27 years of daily data) to test the post-processed calibrated model, we should expect that the predicted quantiles of streamflow would be exceeded on average (leaving aside considerations of elicitability of statistical features of distribution tails, discussed later in Section 4) once in those 10 000 days.

The estimated EVI (see Section 2.3) is also of interest, as it characterizes the tails of the distribution of the simulations, with larger EVI corresponding to heavier tails. The analysis (conducted for all the basins examined in the study) indicates that the estimated EVI depends on the hydrological model and varies from basin to basin (see Figure 3). The medians of the estimated EVIs are equal to 0.19 for the GR4J model, 0.20 for the GR5J model and 0.22 for the GR6J model.

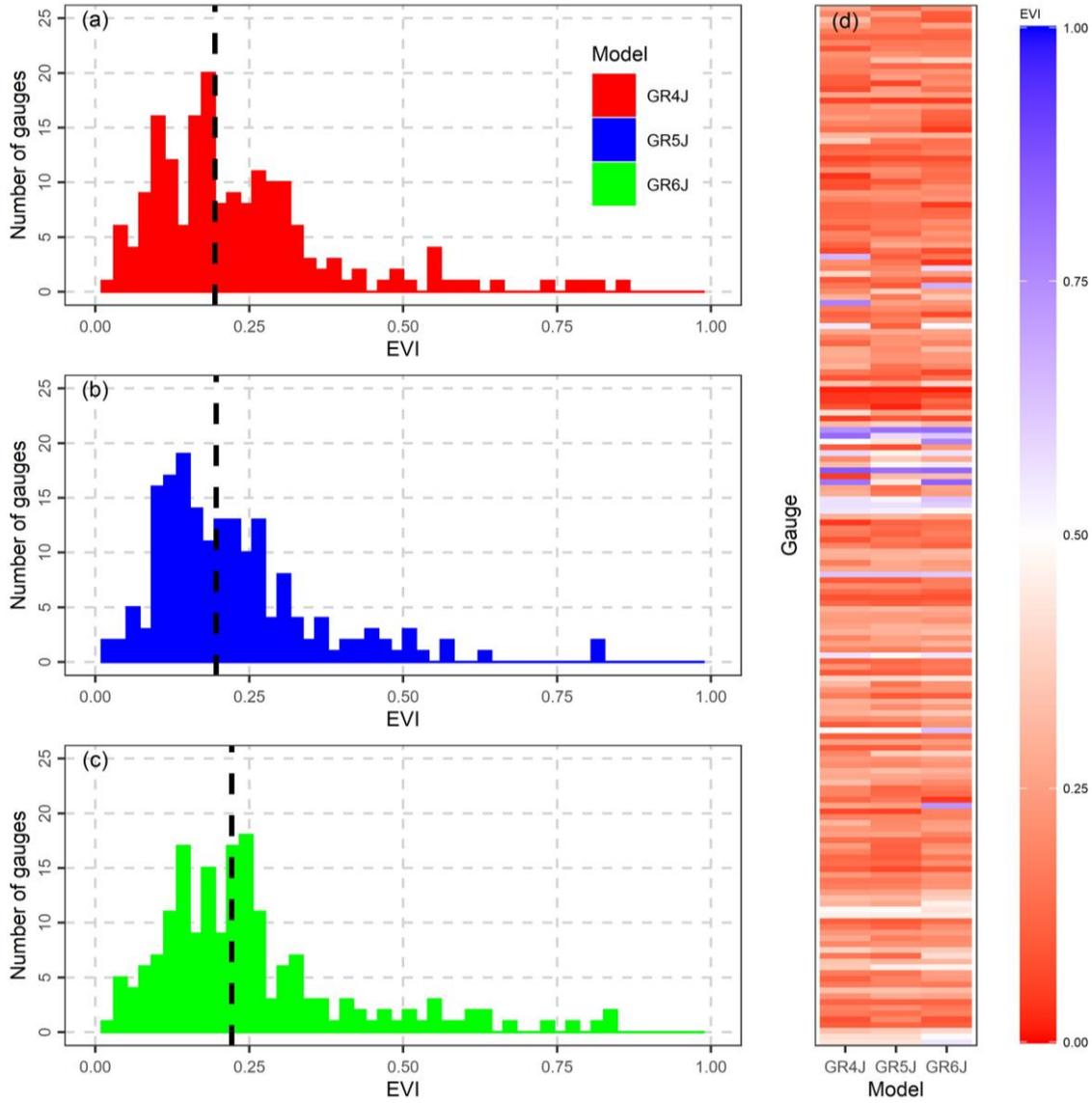

Figure 3. Histograms of the estimated extreme value index (EVI) (the $\gamma$ estimates are obtained by eq. (13)) of the post-processed simulations of the (a) GR4J, (b) GR5J and (c) GR6J models for all the basins, with the post-processing made using extremal quantile regression, and (d) the respective heatmap summarizing results for all the basins. In the histograms, the medians of the estimated EVIs are denoted with thick black dotted lines.

Examining how extremal quantile regression differs from quantile regression is of interest also here. At the level 0.9700, both hydrological post-processing methods are approximately equivalent, with the mean of the logarithms of the ratios of the quantiles estimated by extremal quantile regression to the quantiles estimated by quantile regression, computed across all the examined basins, being equal to 0.00 for GR4J, –0.01 for GR5J and –0.02 for GR6J. However, the quantiles estimated by extremal quantile regression are considerably larger at the level 0.9999. For the same quantile level, the estimated means of the logarithms of the ratios are 0.43 for GR4J, 0.40 for GR5J and 0.47 for GR4J (see Figure 4). As shown in Figure 4, the behaviour of the ratios is consistent during the entire time period at each basin, as it is indicated by the fact that the colouring remains stable when moving on the x-axis.

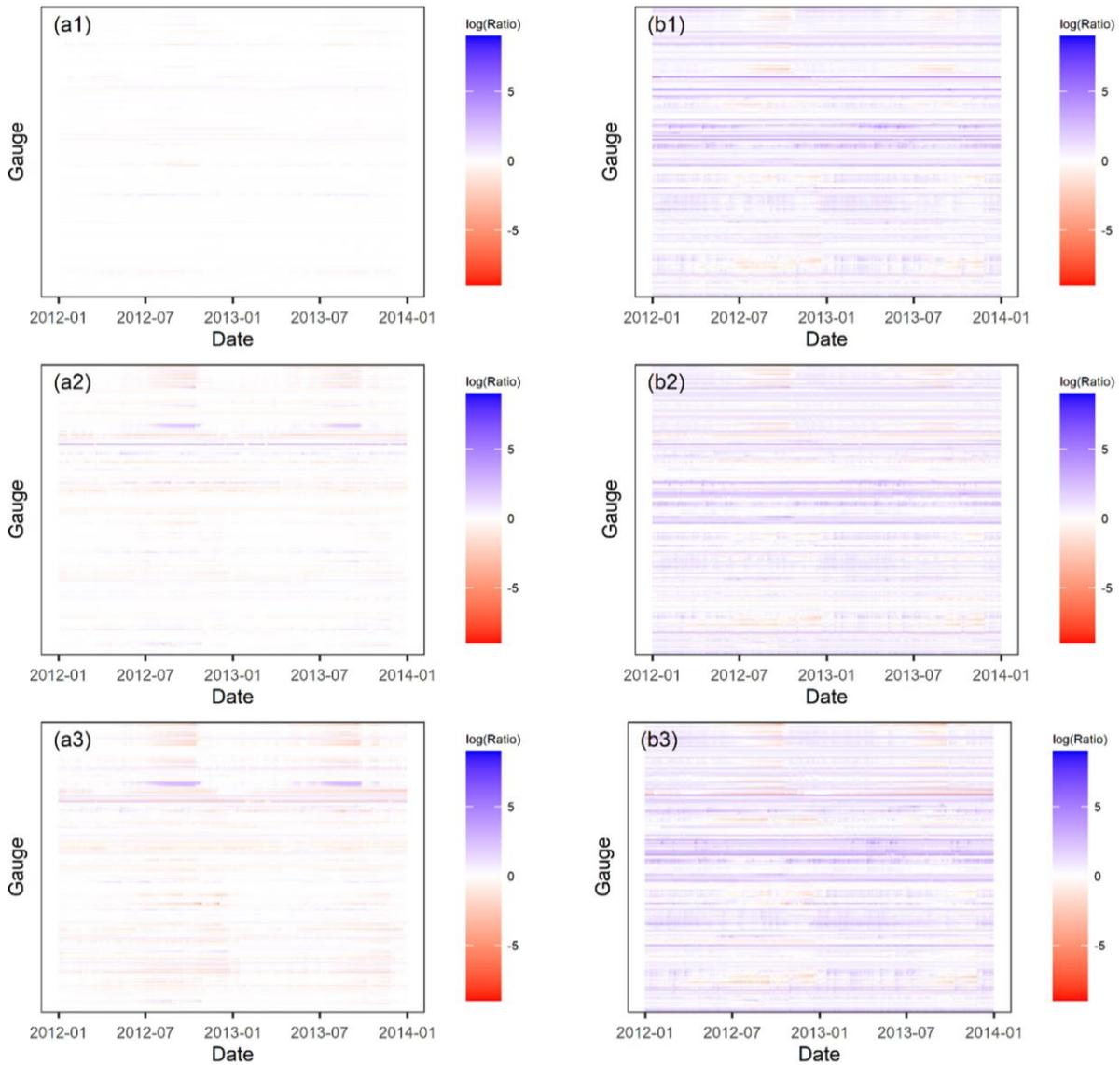

Figure 4. Logarithm of the ratios of the quantiles estimated by the extremal quantile regression to the quantiles estimated by the quantile regression at the levels (a) 0.9700 and (b) 0.9999 when post-processing the simulations by the (a1, b1) GR4J, (a2, b2) GR5J, and (a3, b3) GR6J models at all the basins for the period 2012-2013.

Quantiles estimated by extremal quantile regression at the level 0.9999 seem to be similar for all the three hydrological models, although deviations exist from basin to basin (see Figure 5). The mean of the logarithms of the ratios of the quantiles estimated by post-processing the simulations of the GR5J (GR6J) model to the quantiles estimated by post-processing the simulations of the GR4J model, computed across all the examined basins, equals –0.02 (0.01).

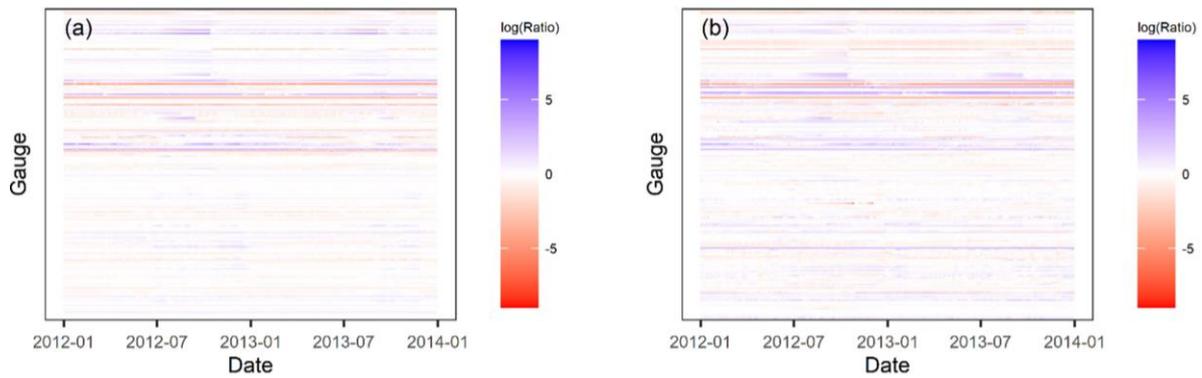

Figure 5. Logarithm of the ratios of the quantiles estimated by post-processing (a) the simulations of the GR5J model or (b) the simulations of the GR6J model to the quantiles estimated by post-processing the simulations of the GR4J model at the level 0.9999 for all the basins and the period 2012-2013.

## 4. Discussion

Extremal quantile regression assumes that the EVI $\gamma$ is higher than 0. Expectantly, that is a befitting assumption for streamflow in CONUS, for which a big data analysis has shown that block annual daily streamflow extremes can be modelled by a Generalized Extreme Value (GEV) distribution with $\gamma > 0$ in the majority of the basins (Tyralis et al. 2019b). Obviously, whereas quantile regression is distribution free (Waldmann 2018), the additional distributional assumptions co-synthesizing extremal quantile regression make it stricter in practice; however, these same assumptions are also inevitable, since they are the ones that offer the possibility for extrapolations beyond the range of the observations. Interpretation of the results considering the attributes of the basins has already been done in Tyralis et al. (2019), in the sense that $\gamma$ of the GEV has already been related to those attributes, while one should expect that similar behaviour will be observed for the herein estimated EVI using the Hill's estimator.

Prediction of extremes has been the focus of intense research, including among others forecasting competitions (Park et al. 2018), special issues (Seneviratne and Zwiers 2015) and regional or even global forecasting services (Yuan et al. 2015) in many and diverse environmental fields. For instance, Becker et al. (2013) have used anomaly correlation and RMSE to assess the skills of models in predicting climate extremes. In another related example on simulated time series, the Brier score has been proposed and adopted for assessing the performance of methods (Williams et al. 2014). Modifying the objective function by using thresholds is also an idea conceived and exploited for predicting extremes (Franch et al. 2020), while modelling using probability distributions is another idea conceived and adopted in the same endeavour (Li et al. 2021b). However, assessing the performance of such methods can also get troublesome, given the largely uncharted nature of the investigated problem, especially in applied fields but in theoretical fields as well. Split-sample techniques are known to be appropriate for assessing the performance of hydrological models (Klemeš 1986, Biondi et al. 2012). Still, the assessment of predictions of extremes could not be formulated as a straightforward process (Lerch et al. 2017), as it has also been attentively discussed by Huser (2021) in the important context of forecasting competitions.

Focusing on hydrological post-processing and motivated by the previous commentary, using quantile scores (Gneiting 2011) for assessing extreme quantile post-processed predictions would not be a good practice, leading to unnatural results. A relevant example (that could be the case as well for the investigations of this work, if our methods had been assessed using quantile scores) would be to erroneously find that quantile regression outperforms extremal quantile regression at high quantiles (at the quantile level 0.9999). The reason behind this specific inconsistency for the investigated problem is that statistical features of distribution tails cannot be elicitable and proper scoring rules (e.g. quantile scores, intervals scores, continuously ranked probability score (CRPS)) cannot separate max-functional values (Brehmer and Strokorb 2019). Additional discussions on scores for extremes and on how to assess point forecasts in absence of established directives can be found in Juutilainen et al. (2012), Schmidt et al. (2021), Taggart (2021) and Martin et al. (2022).

Regarding the results of our paper, they are not comparable to those of other hydrological post-processing studies, due to the unique nature of the examined problem. In particular, conventional quantile regression (also used here), has been shown that cannot extrapolate beyond some upper bounds, therefore being inappropriate to use for extreme events, when the examined time series are short. Results remain stable and mostly independent of the implemented hydrological model, while they also change slightly in time at a given basin. A last remark to be made, before concluding the paper, regards the comparison of the herein proposed approach with previous methods on a theoretical level. An alternative would be ensemble methods (Troin et al. 2021), which are characterized by a low number of ensembles (with a typical number for them being 50), thereby corresponding to lower levels compared to the extreme quantile level of 0.9999 that was examined herein. Parametric models with parameters linked directly to the covariates can also be used, but from our experience such models could be unstable. On the contrary, a single covariate parameter (i.e. the tail index) needs to be estimated by the herein proposed hydrological post-processing method, while the model for the tail is built upon the quantile regression algorithm, borrowing in this way properties from the latter algorithm (Wang et al. 2012). Obviously, all the methods might be unstable when estimating extreme quantiles, with that being especially relevant to the case of short time series, which also applies to the investigations of this paper.

## 5. Conclusions

In this work, extremal quantile regression was proposed for estimating high conditional quantiles of hydrological predictions in the context of uncertainty estimation using post-processing techniques. This practice sets the problem of predicting flood events in a different context, compared to the frequent practice of focusing on the accurate estimation of high flows calibrating the hydrological model using squared error (or similar) loss functions, that is consistent for the task of estimating the mean of the probability distribution of the hydrological model's response (but not the extreme quantiles of the probability distribution that are naturally connected with the concept of floods). The new method successfully post-processed the daily streamflow predictions provided by three process-based hydrological models for 180 basins in the contiguous United States and was compared to the state-of-the-art technique of quantile regression. Under this experimental setting, quantile regression underestimates severely high conditional quantiles, especially when the available data are limited. The equivalence between quantile regression and extreme conditional quantile regression at lower quantile levels was also demonstrated.

In more practical grounds, the use of extremal quantile regression for post-processing daily streamflow predictions leads, on average, to equivalent predictions of high quantiles for the three selected hydrological models, although the results varied from basin to basin. On the other hand, estimates of extremal quantile regression were from 1.5 to 2 times higher, on average, compared to those of quantile regression.

**Conflicts of interest:** The authors declare no conflict of interest.

**Author contributions:** HT and GP contributed equally to this work.

**Acknowledgements:** We thank the Associate Editor and the reviewers whose comments helped us improve the manuscript.

# Appendix A    Open research

The computations and visualizations were conducted in `R` Programming Language (R Core Team 2022) using the following packages: `airGR` (Coron et al. 2017, 2021), `data.table` (Dowle and Srinivasan 2021), `devtools` (Wickham et al. 2021), `EXRQ` (Wang 2016), `gdata` (Warnes et al. 2017), `knitr` (Xie 2014, 2015, 2021), `quantreg` (Koenker 2022), `rmarkdown` (Allaire et al. 2021), `stringi` (Gagolewski 2021), `tidyverse` (Wickham 2021, Wickham et al. 2019). The data are available in Newman et al. (2014) and Addor et al. (2017a).